\documentclass[twocolumn,showpacs,preprintnumbers,amsmath,amssymb]{revtex4}

\usepackage[dvipdfmx]{graphicx}
\usepackage{dcolumn}
\usepackage{bm}
\usepackage{color}
\usepackage{bbm}

\newcommand{\ignoreInFinal}[1]{}

\newcommand{\dt}{\mathrm{dt}}
\newcommand{\Dt}{\Delta t}
\newcommand{\dd}{\mathrm{d}}
\newcommand{\Vc}[1]{\boldsymbol{#1}}
\newcommand{\Prob}{\mathbb{P}}

\newcommand{\Z}{z}
\newcommand{\on}{O}
\newcommand{\off}{\emptyset}

\newcommand{\dW}{\mathrm{d}W}
\newcommand{\diag}{\mathrm{diag}}
\newcommand{\defeq}{:=}
\newcommand{\Transpose}{\mathbb{T}}

\newcommand{\B}{\mathcal{B}}
\newcommand{\R}{\mathcal{R}}
\newcommand{\Erate}{\mathcal{E}}

\renewcommand{\on}{\mathrm{on}}
\renewcommand{\off}{\mathrm{off}}

\newcommand{\seq}{\overset{\circ}{=}}

\newcommand{\ignore}[1]{}

\newcommand{\Input}{U}
\newcommand{\Output}{I}

\begin{document}

\preprint{APS/123-QED}

\title{Bayesian Gates for Reliable Logical Operations under Noisy Condition}

\author{Tetsuya J. Kobayashi }
 \email{tetsuya@mail.crmind.net}
 \homepage{http://research.crmind.net/}
\affiliation{%
Institute of Industrial Science, the University of Tokyo, 4-6-1 Komaba Meguro-ku, Tokyo 153-8505, Japan.
}%

\date{\today}

\begin{abstract}
The reliability of logical operations is indispensable for the reliable operation of computational systems.
Since the down-sizing of micro-fabrication generates non-negligible noise in these systems,  a new approach for designing noise-immune gates is required.
In this paper, we demonstrate that noise-immune gates can be designed by combining Bayesian inference theory with the idea of computation over noisy signals. 
To reveal their practical advantages, the performance of these gates is evaluated in comparison with a stochastic resonance-based gate proposed previously.
We also demonstrate that this approach for computation can be better than a conventional one that conducts information transmission and computation separately.
\end{abstract}

\pacs{Valid PACS appear here}
\maketitle


\ignoreInFinal{\section{Introduction}}
Reliability of logical operations is an indispensable prerequisite for the operation of almost all computations.
Because of the current demand for down-sizing and low energy consumption of computational devices, new designs of logical operations with higher noise-immunity are required. 
To address this problem, biological systems are regarded as good role models, because our body and brain can conduct certain computations robustly and stably, even though their elementary processes, i.e., intracellular reactions and single-neuron spikes, consume low amounts of energy and thereby operate stochastically \cite{Eldar:2010kk,Faisal:2008cp}.

Historically, biology has in fact inspired new designs of noise immune systems.
A Schmitt trigger is an early example, where the study of squid nerves directly led to the idea to use hysteresis for noise-immunity \cite{Anonymous:atfSfZB6}.
More recently, a new implementation of noise-immune logical operations was proposed based on stochastic resonance (SR) \cite{Murali:2009gi,Murali:2009kc,Sinha:2009de},
which has been observed in neural sensory systems to amplify signals with the aid of noise.
Nonetheless, neither hysteresis nor SR is suffices to explain all the noise-immune properties of biological systems. 

A new possible mechanism in the current spotlight is an exploitation of Bayesian computation.
Recent psychological and molecular-biological studies indicated that biological systems ubiquitously employ Bayesian logic for computations under noise and uncertainty from cognitive down to molecular level \cite{Knill:2004ch,Kobayashi:2010vo,Kobayashi:2011ti}. 
This fact suggests that the employment of the Bayesian logic for computation can contribute not only to high-level algorithms but also to low-level gate-design.

In this paper, we demonstrate that Bayesian logic can in fact be utilized for designing noise-immune logic gates in combination with the idea of computation over noisy channels presented in \cite{Nazer:2007cy}.
Bayesian logic gates are shown to have several advantageous properties over the gates based on the previously proposed logical SR (LSR).
In addition, when the noise-level is sufficiently high, a scheme that operates computation over noisy channels (Fig. 1 (A)) can be better than a conventional one where information transmission and computation are conducted separately (Fig. 1 (B)).
Finally, the generality and possible extensions of this approach are discussed.

Let $x_{1}(t) \in \{0, 1\}$ and $x_{2}(t) \in \{0, 1\}$ be two noiseless logical inputs to a logic gate at time $t$. 
The logical input $\Vc{x}(t)=(x_{1}(t), x_{2}(t))^{\Transpose} \in \{0,1\}^{2}$ is generally implemented by a physical state such as voltage as $\Input_{i}(t) = \mu_{i}(\Vc{x}(t)) \in \mathbb{R}$ for $i \in \{1, 2\}$. 
If noise in the physical inputs $\Vc{\Input}(t)\defeq (\Input^{1}(t), \Input^{2}(t))^{\Transpose} \in \mathbb{R}^{2}$ is negligible, a logical operation over $\Vc{x}(t)$, e.g., AND operation $x_{1}(t)\bigwedge x_{2}(t)$,  can be implemented by a two-state thresholded dynamics 
in which the state flips only when both $\Input_{1}(t)$ and $\Input_{2}(t)$ exceed certain threshold values. 
However, when the noise in $\Vc{\Input}(t) $ is sufficiently strong, such dynamics leads to erroneous switching driven by the noise.

The influence of noise in  $\Input_{i}(t)$  is abstractly modeled in this work by the white Gaussian noise as 
\[
\Input_{i}(t)\dt  = \mu_{i}(\Vc{x}(t))\dt + \sigma_{i} \dW^{i}_{t}, \qquad i \in \{1, 2\}.
\]
Here, $\sigma_{i} W^{i}_{t}$ is the one-dimensional Wiener process that represents noise with intensity $\sigma_{i}>0$.
We also assume that $\Input_{i}(t)$ depends only on $x_{i}(t)$, i.e., $\mu_{i}(\Vc{x}(t))=\mu_{i}(x_{i}(t))$.
When the signal-to-noise ratio (SNR) $\Delta \mu_{i}/\sigma_{i}$ is not sufficiently high, where $\Delta \mu_{i}\defeq |\mu_{i}(1)-\mu_{i}(0)|$, the  simple threshold-based switching fails to return the correct output of, for example, the AND operation, because the noisy $\Input_{1}(t)$ and $\Input_{2}(t)$ can exceed the thresholds, even though  $x_{1}(t)=1$ and $x_{2}(t)=1$ do not hold.
This fact illustrates that a simple threshold-based switching does not suffice to implement a reliable logical operation under noise.
To overcome this problem,  we need a dynamical implementation of the logical operations that is more reliable than the simple switching dynamics.

To theoretically derive such an implementation, in this work, we reformulate the logical operation under noise as a statistical inference of partial information. 
In the conventional statistical inference and logic operations, we infer  all hidden states of $\Vc{x}(t)$ from $\Vc{\Input}(t)$ as $\Vc{z}(t)=(\Vc{z}_{1}(t),\Vc{z}_{2}(t))$, where $\Vc{z}(t)$ is the inferred version of $\Vc{x}(t)$.
After inference of transmitted information, the functions of $\Vc{x}(t)$ are calculated with $\Vc{z}(t)$ under noiseless conditions (Fig. 1 (B)). 
However,  the approach here differs from the usual statistical inference in that the main purpose is the inference of partial information of $\Vc{x}(t)$, i.e., a function of $\Vc{x}(t)$, rather than the entire information of $\Vc{x}(t)$, because a logical operation constitutes a reduction of the information on $\Vc{x}(t)$ that $\Vc{\Input}(t)$ posseses. 
This property enables us to conduct the necessary computation over noisy signal $\Vc{\Input}(t)$ before inference, as shown in Fig. 1 (A).
For example, $\Output(t)=\Input_{1}(t) + \Input_{2}(t)$  conveys sufficient information to obtain AND, NAND, OR, NOR, and XOR operations.
This operation over noisy $\Vc{\Input}$ substantially reduces the complexity of computation after inference by avoiding the calculation of the desired output of a gate from the inferred states of $\Vc{x}(t)$. 
Therefore, the concept of computation over noisy signals (channels) is suitable for implementing a reliable circuit by combining unreliable and reliable components effectively, and may also be relevant for biological systems.

\begin{figure}
 \includegraphics[width=0.99\linewidth,bb= 2 21 519 446]{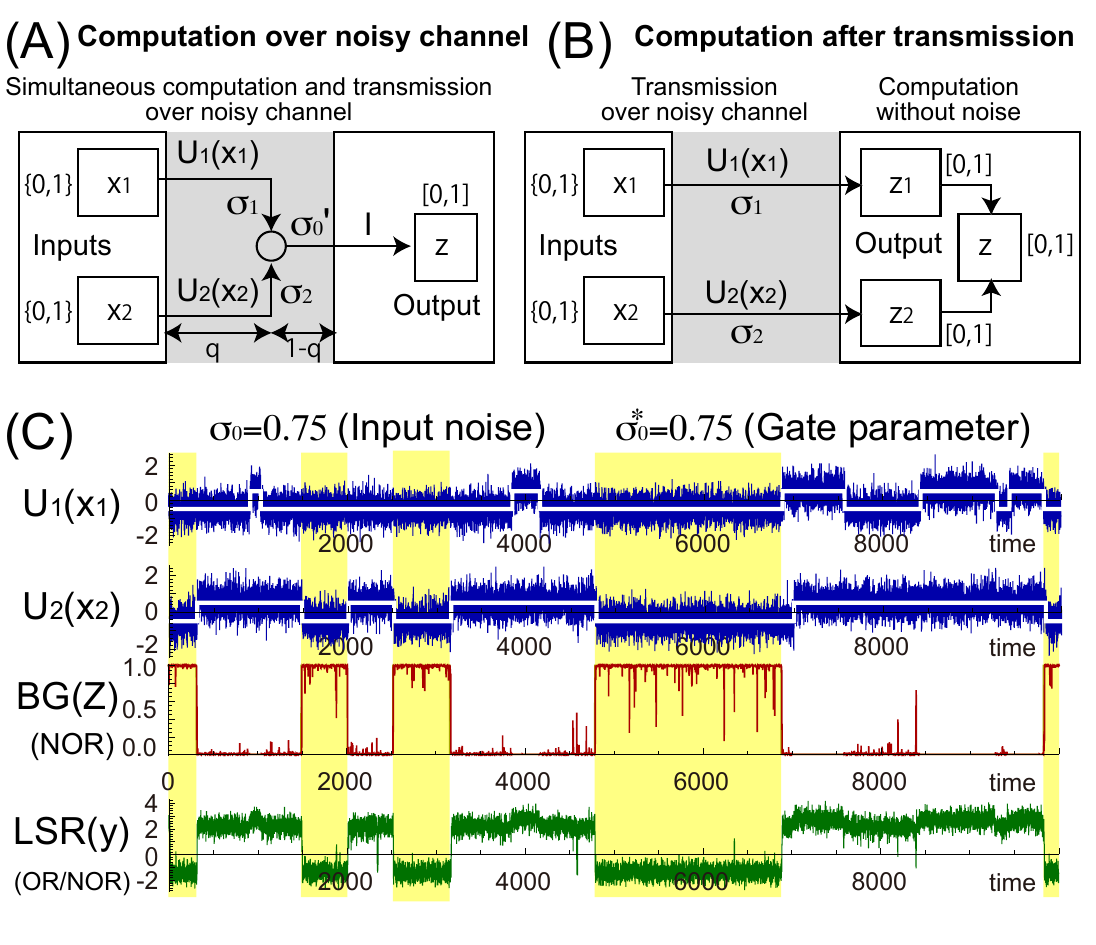}
\caption{\label{fig1} 
Schematic diagram of  Bayesian logic gate (A) and the conventional noiseless computation after information transmission (B).
(C) Sample paths of the noisy inputs $\Input_{1}(t)$ and $\Input_{2}(t)$(blue curves), and the output $\Z_{1}(t)$ of the Bayesian NOR gates and the output $y(t)$ of the OR/NOR gate by the LSR.
The time intervals within which the errorless gate outputs should be $1$ are represented by the filled yellow regions.
The parameters are $\mu=1$, $\sigma_{0}=\sigma_{0}^{*}=0.75$, $r_{\on}=r_{\off}=1/1000$, $y_{l}^{*}=-0.5$, $y_{u}^{*}=1.3$, $\alpha=1.8$, and $\beta=3$.}
\end{figure}

To demonstrate this idea, in this paper, we consider only $\Output(t)=\Input_{1}(t) + \Input_{2}(t)$ as computation over noisy $\Vc{\Input}(t)$, although this approach is applicable for more general situations.
From the definition of $\Input_{i}(t)$ and the properties of the Wiener process, $\Output(t)$ can be simplified as 
$\Output(t)\dt = \nu(\Vc{x}_{t})\dt + \sigma_{0}  \dW_{t}$, where $\nu(\Vc{x}(t))=\mu_{1}(x_{1}(t)) +  \mu_{2}(x_{2}(t))$.
The noise intensity $\sigma_{0}$ depends on the physical implementation of the gate.
For the worst case, where $\Input_{1}$ and $\Input_{2}$ add up just before the inference computation, the noise both in $\Input_{1}$ and $\Input_{2}$ contributes to $\sigma_{0}$ as $\sigma_{0}^{2} = \sigma_{1}^{2}+\sigma_{2}^{2}$. 
In contrast, for the best case,  the noise of the single channel that transmits $\Output(t)$ contributes to the noise as 
$\sigma_{0}^{2} = {\sigma'}_{0}^{2}$.
In order to account for these two extreme situations, we introduce a parameter $q\in[0,1]$ such that $\sigma_{0}^{2} = q(\sigma_{1}^{2}+\sigma_{2}^{2}) + (1-q){\sigma'}_{0}^{2}$ as in Fig. \ref{fig1} (A)  and, for simplicity, assume that $\sigma_{1}=\sigma_{2}={\sigma'}_{0}=\sigma$ to obtain $\sigma_{0}^{2} = (1+ q) \sigma^{2}$.
In addition, we also assume that $\mu_{1}(x)=\mu_{2}(x)=\mu(x)$ because of the symmetry of logic gates with respect to the exchange of two inputs.
Thus, for sufficiently small $\Delta t>0$ and fixed $\Vc{x}(t)$, 
the probability distribution for $\Output(t)$ can be represented as 
$\Prob_{N}(\Output(t); \nu(\Vc{x}(t)), \sigma_{0}^{2}/\sqrt{\Dt})$, where $\Prob_{N}(\Output; \nu, \sigma^{2})$ is the normal distribution, the mean and variance of which are $\nu$ and $\sigma^{2}$, respectively.
Because of its definition, $\nu(\Vc{x}_{t})$ is either $2 \mu(1)$, $\mu(1)+\mu(0)$, or $2 \mu(0)$. 
Thus, $\Output(t)$ can discriminate three of the four possible states of $\Vc{x}_{t}$. 
We designate the three states by $\chi_{i}$ as $\chi_{1}=(0,0)$, $\chi_{2}=(0,1)$ or $(1,0)$, and $\chi_{3}=(1,1)$.
Furthermore, without losing generality, we assume that $\mu(1)=\mu/2$ and $\mu(0)=-\mu/2$.
Now, the inference of $\Vc{x}(t)$ from the noisy input $\Output(t)$ is reduced to the problem of determining  whether $\Vc{x}_{t}$ is in any of $\chi_{i}$.
If we infer whether $\Vc{x}(t)$ is in $\chi_{3}$ or not, then the inference is equivalent to the AND operation, because $\Vc{x}(t)=\chi_{3}$ only when $\Vc{x}(t)=(1,1)$. 
Similarly, we can construct NOR and XOR.
 
The statistically optimal inference of $\Vc{x}_{t}$ is derived by using the sequential Bayesian inference as in \cite{Kobayashi:2010vo}. 
Let $\Z_{i}(t)\defeq \Prob_{t}(\Vc{x}(t)=\chi_{i}| \Output(0:t))$ be the posterior probability that $\Vc{x}(t)=\chi_{i}$ given the history of $\Output(t')$ from time $t'=0$ to $t'=t$. 
By following sequential Bayes' theorem\cite{Berger:1985vo}, we have 
\begin{eqnarray*}
 \Z_{i}(t') \propto \Prob_{N}\left(\Output(t'); \nu(\chi_{i}), \frac{\sigma_{0}^{2}}{\sqrt{\Dt}}\right)   \sum_{j} \Prob_{T}(t',\chi_{i}|t,\chi_{j}) \Z_{j}(t) 
\end{eqnarray*}
where $t'=t+\Dt$, and $\Prob_{T}(t',x_{i}|t,x_{j})$ is the transition probability that $\Vc{x}(t')$ becomes $\chi_{i}$ when $\Vc{x}(t)=\chi_{j}$. 
Then, we have
\begin{eqnarray*}
\frac{\Z_{i}(t') }{\Z_{j}(t')} = \frac{\Prob_{N}(\Output(t'); \nu(\chi_{i}), \frac{\sigma_{0}^{2}}{\sqrt{\Dt}}) }{\Prob_{N}(\Output(t'); \nu(\chi_{j}), \frac{\sigma_{0}^{2}}{\sqrt{\Dt}}) } \frac{ \sum_{k} \Prob_{T}(t',\chi_{i}|t,\chi_{k}) \Z_{k}(t)}{ \sum_{k} \Prob_{T}(t',\chi_{j}|t, \chi_{k}) \Z_{k}(t)}.
\end{eqnarray*}
For simplicity, we assume that $\Prob_{T}(t', \chi_{i}|t, \chi_{j})$ is time-homogeneous and can be represented for sufficiently small $\Delta t$ by $\Prob_{T}(t', \chi_{i}|t, \chi_{j})= \Delta t \times r_{i|j}  $ for $i \neq j$ and $\Prob_{T}(t', \chi_{i}|t, \chi_{i})= 1- \Delta t \times r_{i|i}$ where $r_{i|j}$ is the instantaneous transition rate from $\chi_{j}$ to $\chi_{i}$ and $ r_{i|i}=\sum_{k \neq i}r_{k|i}$ holds. 
If the dynamics of both $x_{1}(t)$ and $x_{2}(t)$ follow a two-state Markov process whose transition rate from $0$ to $1$ and $1$ to $0$ are $r_{\on}$ and $r_{\off}$, respectively, then we have
\[
\R \defeq(r_{i|j})= \begin{pmatrix}
               - 2 r_{\on} & r_{\off}   &                      0  \\
               2 r_{\on}   & -r_{\on} - r_{\off}  & 2 r_{\off}  \\
               0                & r_{\on}  & -2 r_{\off} 
            \end{pmatrix}.
\]
By taking the limit as $\Delta t \to 0$, we obtain a three dimensional equation with quadratic nonlinearity as
\begin{eqnarray*}
\frac{\dd }{\dt}\left(\log \frac{\Z_{i}(t)}{\Z_{j}(t)}\right) &=& \B_{i,j}(\Output(t))+ \frac{[\R\Vc{\Z}(t)]_{i}}{\Z_{i}(t)}- \frac{[\R\Vc{\Z}(t)]_{j}}{\Z_{j}(t)},
\end{eqnarray*}
where $\Vc{\Z}(t)=(\Z_{1}(t),\Z_{2}(t),\Z_{3}(t))^{\Transpose}$ and
\begin{eqnarray*}
\B(\Output(t)) &=& \frac{\mu}{\sigma_{0}^{2}}\left[
	    \Output(t)\begin{pmatrix}
               0  & -1  & -2 \\
              1  & 0  & -1   \\
              2 & 1  & 0 
            \end{pmatrix} 
            + \frac{\mu}{2}
            \begin{pmatrix}
               0  & -1  & 0  \\
              1  & 0  & 1   \\
              0 & -1  & 0 
            \end{pmatrix}\right], \\
\end{eqnarray*}
Finally, by solving the above equation with respect to $\Vc{\Z}(t)$,  we have 
\begin{eqnarray}
\frac{\dd \Vc{\Z}(t)}{\dt} &\seq& \diag(\Vc{\Z}(t))  \B(\Output(t))\Vc{\Z}(t) + \R\Vc{\Z}(t), \label{eq:BGA}
\end{eqnarray}
where $\seq$ indicates that the integrals with respect to $\dW_{t}$ are interpreted as the Stratonovich integrals \cite{Gardiner:1985tb}( see supplementary material for the detailed derivation).
The three outputs, $\Z_{1}(t)$, $\Z_{2}(t)$, and $\Z_{3}(t)$, correspond to Bayesian NOR, XOR, and AND gates, respectively, 
and therefore, this system can simultaneously compute these operations.
By an appropriate transformation of variables, we can also implement OR, NOR, and NAND operations.
Equation (\ref{eq:BGA}) contains $\sigma_{0}$ and $\mu$ as the system's parameters, indicating that the optimal tuning of these parameters such that they coincide with those of the input $\Output(t)$ is necessary to realize statistically optimal logical operations. 
However, the gate parameters may not be accurately adjusted in a real situation. 
In order to analyze the influence of such a parameter mismatch, we introduce $\mu^{*}_{0}$ and $\sigma^{*}_{0}$ to specifically represent the gate's parameters $\mu$ and $\sigma_{0}$ in Eq. \eqref{eq:BGA}, and therefore, Eq. \eqref{eq:BGA} is statistically optimal only when $\mu^{*}_{0}=\mu$ and $\sigma^{*}_{0}=\sigma_{0}$.

\ignoreInFinal{\section{Performance of Bayesian gates }}
Figures 1(C) and S1(A) demonstrate that the Bayesian gates can conduct logical operations under a very  noisy condition. 
We compare the performance of the Bayesian gates with LSR proposed recently as a noise-immune implementation of logical gates \cite{Murali:2009gi,Murali:2009kc}.
LSR is defined by the following stochastic differential equation with a double-well potential:
\[
\dd y = \left[-\alpha y + \beta g(y) + \Output(t)\right]\dt,
\]
where $g(y) = y$ when $y_{l}^{*} \le y \le y_{u}^{*}$, $g(y)=y_{l}^{*}$ when $y <y_{l}^{*}$, and $g(x)=y_{u}^{*}$ when $y >y_{u}^{*}$.
The optimal LSR NOR gate can be implemented by setting $(y_{l}^{*}, y_{u}^{*})=(-0.5, 1.3)$ as in \cite{Murali:2009gi}.
In order to evaluate the performance of the Bayesian and LSR NOR gates, 
we use an error rate (ER) defined as $\Erate \defeq \frac{1}{T} \int_{0}^{T} \left| \Vc{1}[a(t)>a_{\mathrm{th}}] - G[x_{1}(t);x_{2}(t)] \right| \dt$, where $a(t)$ is either $\Z_{i}(t)$ or $y(t)$.
$\Vc{1}[a>a_{\mathrm{th}}]$ returns $1$ when $a>a_{\mathrm{th}}$ and $0$ otherwise.
For $a(t)=\Z_{i}(t)$, we choose $a_{\mathrm{th}}=1/2$, whereas we choose $a_{\mathrm{th}}=0$ for $a(t)=y(t)$ as in \cite{Murali:2009gi}.
$G[x_{1};x_{2}]$ is the output of the ideal noiseless gate with input $\Vc{x}$.
If the NOR gate is concerned, for example, $G[x_{1};x_{2}]$ returns $1$ when $(x_{1},x_{2})=(0,0)$, and $0$ otherwise.
The performance of LSR is shown to be optimal when $0.6< \sigma_{s}<0.8$, as in Fig. 2(A).
Under this optimal noise intensity for the LSR gate, the ERs of the LSR and the Bayesian gates are comparative, indicating that the performance of the LSR gate is close to the statistical  optimal (Figs. 1(C) and 2(A)).

\begin{figure}
 \includegraphics[width=0.99\linewidth,clip]{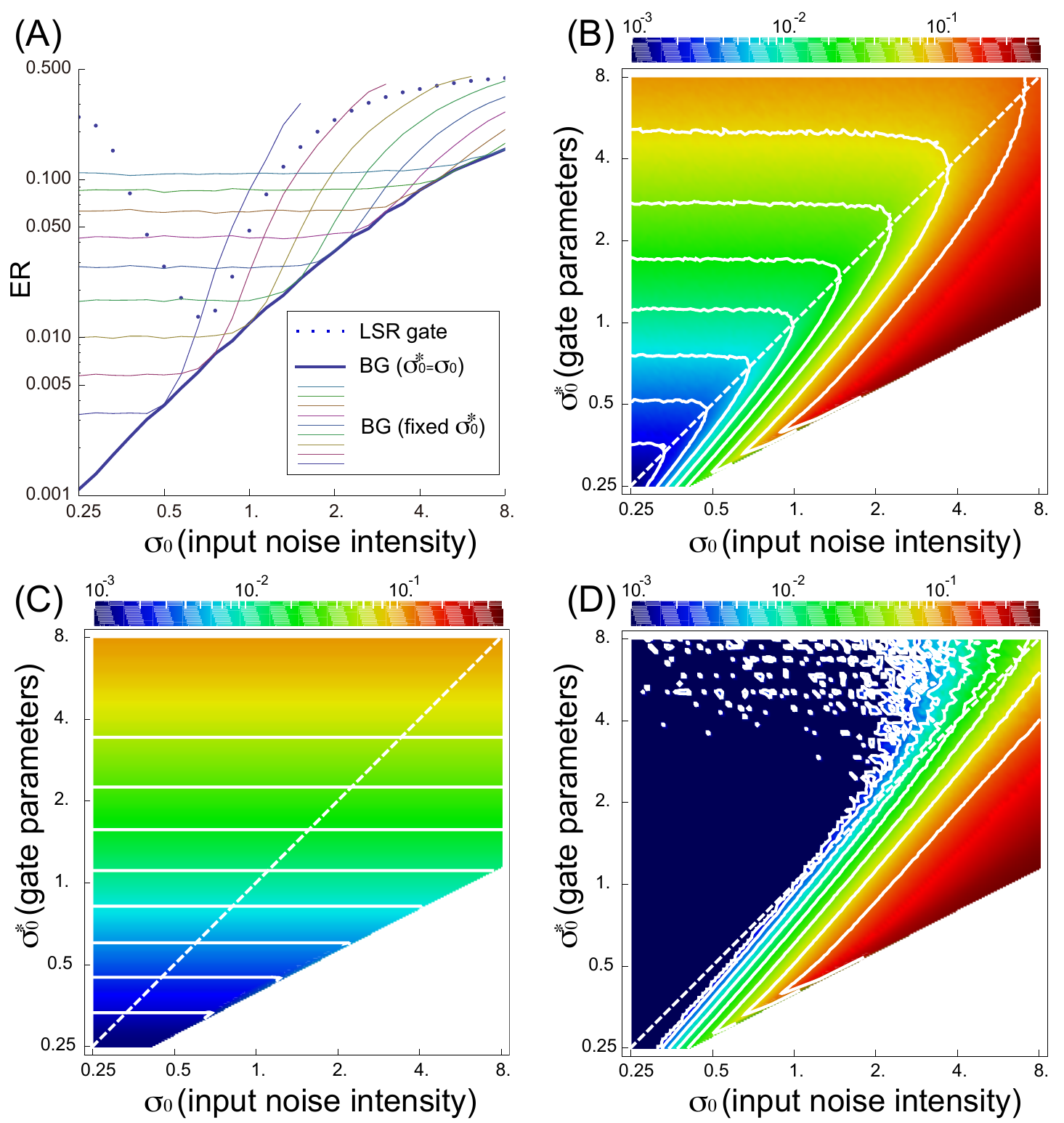}
\caption{\label{fig2}
(A) Error rates $\Erate$ of the Bayesian NOR (solid curves) and the LSR NOR (dotted curve) gates as a function of the noise intensity $\sigma$ of the input. 
Thin colored curves correspond to the error rate of the Bayesian NOR gate with different $\sigma^{*}_{0}$. 
The thick blue curve represents the error rate of the Bayesian NOR gate whose the parameter $\sigma^{*}_{0}$ is adjusted to be optimal as $\sigma^{*}_{0}=\sigma_{0}$.
(B) Total error rate  $\Erate$ of the Bayesian NOR gate as a function of $\sigma_{0}$ and $\sigma^{*}_{0}$. 
The broken white line represents $\sigma_{0}=\sigma^{*}_{0}$. 
The solid white curves are the contours of error rates.
(C) Error rate of the Bayesian NOR gate due to the delay of switching, $\Erate_{D}$. 
(D) Error rate of the Bayesian NOR gate due to erroneous switching by input noise, $\Erate_{E}$. 
The other parameters than $\sigma_{0}$ and $\sigma_{0}^{*}$ are the same as those in Fig. 1.}
\end{figure}

As shown in Figs. 2(A), S1(B), and S1(C), however, the performance of the LSR gate quickly degrades if the noise intensity of the input, $\sigma_{0}$, deviates from its optimal one, whereas the Bayesian gate can still conduct a reliable logical operation within a wider range of noise intensity (Fig. 2(B)).
Furthermore, the ER of the Bayesian gate does not increase if $\sigma_{0}$ is less than the gate parameter, $\sigma^{*}_{0}$.
This property of the Bayesian gate means that its performance is determined by the worst noise level that $\sigma^{*}_{0}$ specifies.
As long as the actual noise intensity $\sigma_{0}$ is less than this expected worst level $\sigma^{*}_{0}$, the gate operates robustly at the cost of a fixed lower bound of the ER.
Since information of the actual noise level within a system may not always be available before designing gates, the Bayesian gate has a practical advantage over the LSR gate.
In general, the error stems from either erroneous switching for constant $x_{t}$ or delay in the switching of $\Z_{t}$ when $x_{t}$ changes.
Since the total ER can be attributed only to the delay of switching when no noise exists in the input as $\sigma_{0} = 0$, 
we can approximately dissect the total ER $\Erate$ into the errors from the delay of switching at the changes of $x_{t}$ as $\Erate_{D}:=\lim_{\sigma_{0} \to 0}\Erate$, (Fig. 2(C)) and those from the erroneous switching for constant $x_{t}$ as $\Erate_{E}:=\Erate - \Erate_{D}$, (Fig. 2(D)). 
As clearly seen in Fig. 2(D), $\Erate_{E}$ hardly changes if $\sigma^{*}_{0}$ (the expected noise intensity) is larger than $\sigma_{0}$ (the actual noise intensity), whereas $\Erate_{D}$ increases.
Thus, the cost of choosing $\sigma^{*}_{0}$ larger than $\sigma_{0}$ is the delay of switching, which limits the speed of the gate.
However, $\sigma^{*}_{0}$ larger than $\sigma_{0}$ works as a margin for systematic variations of $\sigma_{0}$, because $\Erate_{E}$ increases little provided that $\sigma_{0}$ is less than $\sigma^{*}_{0}$.
The same result is obtained for Bayesian XOR and AND gates (Fig. S2).
Thus, the variation in $\sigma_{0}$ can be compensated at the cost of slow switching, reflecting a trade-off between the computational speed and the reliability of computation \cite{Gammaitoni:2007ju}.

\ignoreInFinal{\section{Further Advantage of Computation over Noisy Signal}}
Information transmission and computation are usually separated in conventional computational architectures in which computation is conducted under virtually noiseless conditions (Fig. 1(B)).
This usual computation without noise is expected to perform better than the Bayesian gates that conducts computation and transmission simultaneously in noisy conditions. 
However, it requires two independent channels for input transmission (Fig. 1 (B)), whereas the Bayesian gate combines them along the transmission path for the computation (Fig. 1 (A)).
This observation suggests that the Bayesian gate may outperform the usual computation if a noisy channel is effectively exploited for computation with small $q$.
In order to clarify this condition, we calculated $\Erate(q)$ of the Bayesian NOR gate as a function of $q \in [0,1]$, and that of the usual computation $\Erate_{U}$ to obtain a performance ratio defined as $\eta \defeq \Erate(q)/\Erate_{U}$.
Figure 3(A) shows that the operation of the Bayesian gate can be more efficientl ($\eta < 1$) than or comparative ($\eta < 10^{0.1} \approx 1.25$) to the usual computation when $q$ is sufficiently small.
In addition, the range of $q$ within which the Bayesian gate operates better expands as the noise intensity $\sigma$  increases.
The same result is obtained for other gates (Fig. S3).
This result indicates that the operation of the Bayesian gate can be efficient when the noise in the channel is large, whereas the usual computation is better when the channel noise is very small, 
meaning that the computation over noisy channels may be practical when the noise cannot be small. 

\begin{figure}
 \includegraphics[width=0.75\linewidth,clip]{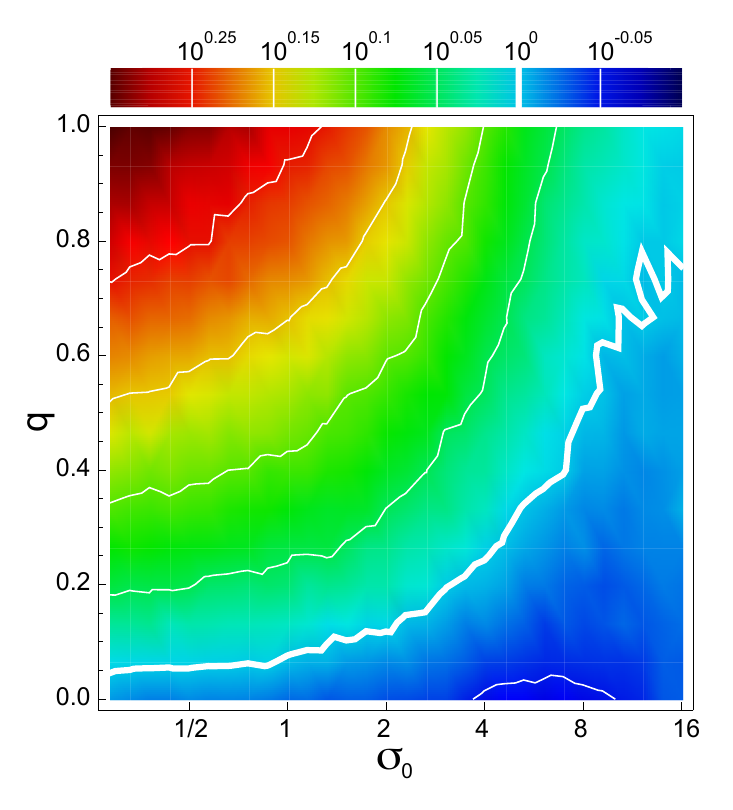}
\caption{\label{fig3} 
The performance ratio $\eta$ of the Bayesian NOR gate as a function of $\sigma_{0}$ and $q$. 
The white curves represent contours of $\eta$ where the thick white curve corresponds to $\eta=1$. 
The gate parameter is set to be optimal as $\sigma_{0}^{*}=\sigma_{0}$.
The other parameters are the same as those in Fig. 1(A).
}
\end{figure}

\ignoreInFinal{\section{Discussion}}
Since our approach is based on the general theory of inference not on a specific physical implementation, it potentially has more extensions and applications than those demonstrated here. 
First, we can choose arithmetic operations other than addition over noisy signals $U_{1}(t)$ and $U_{2}(t)$, which lead to different noise characteristics and gate properties.
For example, subtraction may lead to more reliable gates than addition in principle by canceling out the correlated noise in $U_{1}$ and $U_{2}$. 
Second, we can easily design noise-immune gates with more than two inputs to conduct more complicated logical operations at the cost of the complexity of individual gates.
Third, noise is not restricted to Gaussian white noise.
For example, an equation similar to Eq. 1 can be derived for Poisson noise, which is more relevant to gate implementations by photons or by intracellular reactions. 
Since intracellular systems are known to conduct various computations \cite{Navlakha:2011ik}, 
the Poissonian version may help us to understand and synthetically  design intracellular information processing networks  \cite{Purnick:2009gt}.
The combination of computation over noisy signals and the design of reliable gates by inference theory proposed in this paper is sufficiently general to cover all these situations and should be investigated further for individual situations.


\ignoreInFinal{\begin{acknowledgments}}
We thank Ando, Yuzuru Sato, Atsushi Kamimura, Yoshihiro Morishita, and Ryo Yokota for fruitful discussions. 
This work was supported partially by JSPS KAKENHI Grant Numbers 15H00800 and 19H05799,  the Platform for Dynamic Approaches to Living Systems funded by MEXT and AMED, Japan, and the JST PRESTO program.

\end{document}